\documentclass{article}
\usepackage[T1]{fontenc} %
\usepackage[utf8]{inputenc} %
\usepackage{ismir,amsmath,cite,url}
\usepackage{graphicx}
\usepackage{color}
\usepackage{amsmath,amssymb,graphicx,url,times}
\usepackage{textcomp}
\usepackage{multirow}
\usepackage{etoolbox,siunitx}
\usepackage{pifont}
\usepackage{booktabs}
\robustify\bfseries

\usepackage{lineno}

\title{Sequence-to-Sequence Piano Transcription\\with Transformers}

\multauthor
{Curtis Hawthorne \hspace{1cm} Ian Simon \hspace{1cm} Rigel Swavely} { \bfseries{Ethan Manilow$^{\star}$\thanks{$^{\star}$ Work done as a Google Brain Student Researcher.} \hspace{1cm} Jesse Engel}\\
Google Research\\
{\tt\small \{fjord,iansimon,rigeljs,emanilow,jesseengel\}@google.com}
}

\def\authorname{C. Hawthorne, I. Simon, R. Swavely, E. Manilow, and J. Engel}

\usepackage[bookmarks=false,pdfauthor={\authorname},pdfsubject={\papersubject},hidelinks]{hyperref}

\begin{document}

\maketitle
\begin{abstract}
Automatic Music Transcription has seen significant progress in recent years by training custom deep neural networks on large datasets. However, these models have required extensive domain-specific design of network architectures, input/output representations, and complex decoding schemes. In this work, we show that equivalent performance can be achieved using a generic encoder-decoder Transformer with standard decoding methods. We demonstrate that the model can learn to translate spectrogram inputs directly to MIDI-like output events for several transcription tasks. This sequence-to-sequence approach simplifies transcription by jointly modeling audio features and language-like output dependencies, thus removing the need for task-specific architectures. These results point toward possibilities for creating new Music Information Retrieval models by focusing on dataset creation and labeling rather than custom model design.
\end{abstract}
\section{Introduction}\label{sec:introduction}

\begin{figure*}[t]
    \centering
    \includegraphics[width=\textwidth]{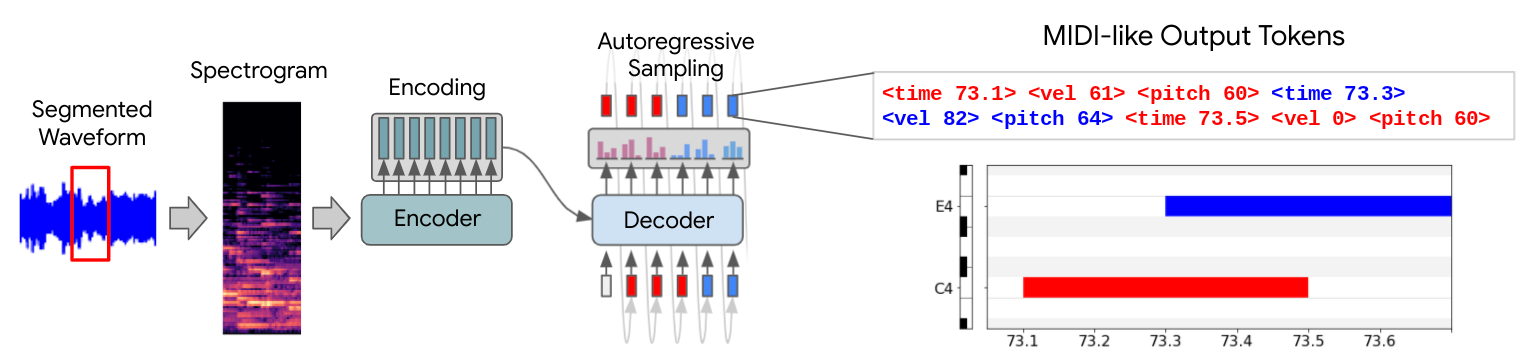}
    \caption{Our model is a generic encoder-decoder Transformer architecture where each input position contains a single spectrogram frame and each output position contains an event from our MIDI-like vocabulary. Outputs tokens are autoregressively sampled from the decoder, at each step taking the token with maximum probability.}
    \label{fig:mirtl_diagram}
\end{figure*}

Automatic Music Transcription (AMT) is one of the core tasks of Music Information Retrieval (MIR).  The objective of AMT is to convert raw audio to a appropriate symbolic representation.  In this paper we consider the problem of transcribing piano audio to a series of note events indicating precise onset/offset timings and velocities, as opposed to a sheet music score that is aligned to a metrical grid.

Recent progress in piano transcription has been largely driven by two factors: the construction and release of datasets containing aligned piano audio and MIDI (most notably MAPS \cite{emiya2009multipitch} and MAESTRO \cite{hawthorne2019enabling}) and the use of deep neural networks with architectures specifically designed for piano transcription (e.g., the Onsets and Frames architecture that models note onsets and note presence separately \cite{hawthorne2017onsets}). While domain-specific models have lead to improvements on benchmark datasets, it is not clear if these approaches can translate to other domains and MIR tasks. 

Simultaneously, Transformer models employing self-attention~\cite{AttentionIsAllYouNeed} have demonstrated a surprising ability to achieve state-of-the-art results in a variety of domains with the same core architecture, by simply varying the input and output representations~\cite{dosovitskiy2021an,2020t5,jaegle2021perceiver, carion2020end, ramesh2021zero, lin2020end, dong2018speech, pham2019very, li2019neural, gong2021ast}.

In this paper, we demonstrate that a generic Transformer model can achieve state-of-the-art piano transcription without any domain-specific adaptations. Using ``off-the-shelf'' components (an essentially unmodified encoder-decoder configuration from the T5 paper~\cite{2020t5}) and a simple greedy decoding strategy, we train a model to encode raw spectrogram frames and decode directly to a sequence of note events inspired by messages in the original MIDI protocol~\cite{midi1996complete} (e.g., \texttt{note on} and \texttt{velocity} messages). As such, we refer to the model's output as ``MIDI-like'' throughout the rest of this paper. We provide details of our model's vocabulary in Section~\ref{subsec:inputsoutputs}.

Further, we demonstrate that this domain-agnostic approach enables us to train several variations on the transcription task (e.g., transcribing only note onsets) by changing only the training labels and without modifying the inputs or model. 

In summary, this work illustrates the value of using generic sequence-to-sequence Transformers for piano transcription, without domain-specific adaptations, and points to the potential to expand a similar approach to a variety of MIR tasks.

\section{Related Work}

\subsection{Piano Transcription}
Much progress has been made in piano transcription using deep neural network models trained on datasets of aligned audio and MIDI.  In 2012, Boulanger-Lewandowski et al. \cite{boulanger2012modeling} (building off the acoustic model of Nam et al. \cite{nam2011classification}) trained a recurrent neural network (RNN) transcription model to output a binary piano roll.  B{\"o}ck and Schedl \cite{bock2012polyphonic} trained a similar RNN-based model for piano onsets only.  Hawthorne et al. \cite{hawthorne2017onsets} improved transcription accuracy by having separate convolution-based model \emph{stacks} for detecting note onsets, note presence, and note velocities. Model outputs were decoded into discrete notes using a hard prior by not initiating a note unless the onset predictor gave probability more than $0.5$.

More recently, progress on piano transcription has largely involved adding more domain-specific deep neural network components and modifying the decoding process.  For the most part, this additional complexity has been geared toward the specific purpose of improving piano transcription accuracy.

Kong et al. \cite{kong2020high} achieve higher transcription accuracy by using regression to predict precise continuous onset/offset times, using a similar network architecture to Hawthorne et al. \cite{hawthorne2017onsets}.  Kim \& Bello \cite{kim2019adversarial} use an adversarial loss on the transcription output to encourage a transcription model to output more plausible piano rolls. Our sequence-to-sequence approach explicitly models such inter-output dependencies through the autoregressive decoder, which is trained end-to-end with the encoder that extracts meaningful audio features.

Kwon et al. \cite{kwon2020polyphonic} use a language model of sorts to model per-pitch note state transitions instead of having separate onset, frame, and offset stacks.  However, the decoding process is fairly complex, in particular the handling of interactions between different pitches.  Similarly, Kelz et al. \cite{kelz2019deep} decode using a hidden Markov model over note states based on attack-decay-sustain-release (ADSR) envelopes.

In a very thorough domain-specific treatment, Elowsson \cite{elowsson2020polyphonic} constructs a hierarchical model that extracts fundamental frequency contours from spectrograms and uses these contours to infer note onsets and offsets.  While it can be useful for many applications to have such intermediate representations as in Engel et al. \cite{engel2020self}, in this work we treat polyphonic transcription from audio to discrete notes as an end-to-end problem.  This has the advantage of conceptual simplicity and our evaluation (\secref{sec:evaluation}) shows it is also effective.

\subsection{Transformers}
Recently, a generic Transformer architecture \cite{AttentionIsAllYouNeed} has been used across multiple domains to solve sequence-to-sequence problems, replacing task-specific architectures that had previously been in use.  Outside the field of natural language processing in which Transformers initially emerged and are now widely used (e.g., GPT-3 by Brown et al. \cite{brown2020language} and T5 by Raffel et al. \cite{2020t5}), Transformers have been used in computer vision for tasks such as object detection \cite{carion2020end}, caption-based image generation \cite{ramesh2021zero}, and pose reconstruction \cite{lin2020end}, as well as audio-related tasks including speech recognition \cite{dong2018speech, pham2019very}, speech synthesis \cite{li2019neural}, and audio event classification \cite{gong2021ast}.

Note that many of the above uses of Transformer take advantage of a \emph{pretraining} phase where the model is trained on a large amount of unlabeled data using self-supervision.  While it is possible that such a pretraining phase might also help with music transcription, in this work we explore the simpler setting of training the Transformer architecture from scratch on labeled transcriptions in an ordinary supervised fashion.

\subsection{Sequence-to-Sequence Transcription}
The idea of using Transformers for music transcription has also been considered.  Awiszus in 2019 \cite{awiszus2019automatic} explored several formulations of music transcription as a sequence-to-sequence problem, using a variety of input and output representations (including ones similar to our own) with both LSTM \cite{hochreiter1997long} and Transformer models.  However, the paper was unable to demonstrate clear success, seemingly due to using a framewise multi-F0 evaluation rather than the note-based evaluation standard in piano transcription, using relative time shifts rather than absolute (see \secref{subsec:inputsoutputs}), and training on the MAPS dataset which is much smaller than the MAESTRO dataset we use.  Earlier, Ullrich and van der Wel \cite{ullrich2017music} appear to be the first to have posed music transcription as a sequence-to-sequence problem (using LSTMs instead of Transformers), but their system could handle only monophonic music.

\section{Model}

As mentioned above, our model is a generic encoder-decoder Transformer architecture where each input position contains a single spectrogram frame and each output position contains an event from our MIDI-like vocabulary. An overview of our model and our input and output setup is shown in Figure~\ref{fig:mirtl_diagram}.

Inputs are processed through a stack of encoder self-attention layers resulting in a sequence of embeddings the same length as the original input.  A stack of decoder layers then uses both causally masked self-attention over the decoder output and cross-attention over the full output of the encoder stack. Crucially, this allows the symbolic token output to be variable length, dependent only on the number of tokens needed to describe the input audio.

\subsection{Model Architecture}

The model configuration is based on the ``small'' model from T5 \cite{2020t5}, with modifications as suggested by the T5.1.1 recipe\footnote{\url{https://github.com/google-research/text-to-text-transfer-transformer/blob/master/released_checkpoints.md\#t511}}. Specifically, our model uses an embedding size of $d_{\mathrm{model}} = 512$, a feed-forward output dimensionality of $d_{\mathrm{ff}} = 1{,}024$, a key/value dimensionality of $d_{\mathrm{kv}} = 64$, $6$-headed attention, and $8$ layers each in the encoder and decoder.

Our model has a few minor changes from the standard configuration.  Most important is that in order to use continuous spectrogram inputs, we add a dense layer to project each spectrogram input frame to the Transformer's input embedding space.  We also use fixed absolute positional embeddings rather than the logarithmically scaled relative positional bucket embeddings used in T5, to ensure all positions can be attended to with equal resolution.  Finally, we use \texttt{float32} activations for better training stability because our model is small enough that we do not need the memory efficiency of the less precise \texttt{bfloat16} format \cite{abadi2016tensorflow} typically used in large T5 models.

The model is implemented using the T5X framework\footnote{\url{https://goo.gle/t5x}}, which is built on Flax \cite{flax2020github} and JAX \cite{jax2018github}.  We also use SeqIO\footnote{\url{http://github.com/google/seqio}} for data preprocessing and evaluation.  Code for our implementation will be available at \url{https://goo.gl/magenta/seq2seq-piano-transcription-code}.

While some recent research using Transformers has favored very large models, such as GPT-3 \cite{brown2020language} with 175B parameters, we found that a comparatively small model is sufficient for these tasks.  With the configuration described above, our model has only 54M parameters, only roughly twice that of Onsets and Frames \cite{hawthorne2019enabling}, which has 28M parameters.

\begin{figure*}[t]
    \centering
    \includegraphics[width=\textwidth]{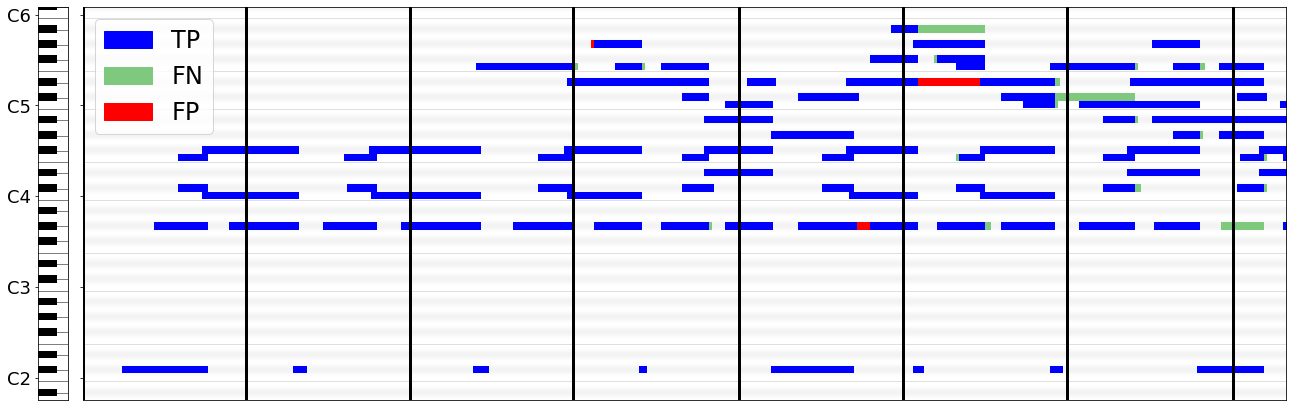}
    \caption{Section of a piano roll rendering of model event output on Chopin's \textit{Berceuse Op.~57 in D-flat Major} from the MAESTRO validation set versus the ground truth.  Black vertical lines represent segment boundaries during inference. True positive (TP) frames are marked in blue, false negatives (FN) in green, and false positives (FP) in red.  Note that the model successfully predicts note-off events for notes where the note-on event happened in a different segment.}
    \label{fig:inference_frames}
\end{figure*}

\subsection{Inputs and Outputs}
\label{subsec:inputsoutputs}

The model uses spectrogram frames as input, with one frame per input position.  To match the T5 setup, we terminate input sequences with a learnable EOS (End of Sequence) embedding.  The model output at each step is a softmax distribution over a discrete vocabulary of events, described below.  This vocabulary is heavily inspired by the messages originally defined in the MIDI specification~\cite{midi1996complete}. Using events as the output representation instead of a piano roll matrix has the advantage of being much more sparse because outputs are needed only when an event occurs, instead of needing to annotate every frame.  The vocabulary consists of the following token types:

\begin{description}
    \item[Note] [128 values] Indicates a note-on or note-off event for one of the 128 MIDI pitches. We use the full MIDI pitch range for flexibility, but for these experiments, only the 88 pitches corresponding to piano keys are actually used.
    \item[Velocity] [128 values] Indicates a velocity change to be applied to all subsequent Note events (until the next Velocity event).  There are 128 velocity values including zero, a special value which causes subsequent Note events to be interpreted as note-off events.
    \item[Time] [6,000 values] Indicates the absolute time location within the segment, quantized into 10 ms bins.  This time will apply to all subsequent Note events until the next Time event.  Time events must occur in chronological order. We define the vocabulary with times up to 60 seconds for flexibility, but because time resets for each segment, in practice we use only the first few hundred events of this type.
    \item[EOS] [1 value] Indicates the end of the sequence.
\end{description}

Previous work that used such a MIDI-like event vocabulary \cite{oore2020time} used relative time shifts between events, indicating the amount of time elapsed since the last time shift.  However, in the sequence-to-sequence scenario a single relative time shift error early in the output causes all subsequent output steps to be incorrect, and such errors accumulate as sequence length increases.  To adjust for this drift, the Transformer model would have to learn to perform a cumulative sum over all previous time shifts in order to determine the current position in time.  We instead use absolute time, where each time event indicates the amount of time from the beginning of the segment, as illustrated in \figref{fig:mirtl_diagram}. This gives the model the easier task of determining each timestamp independently; we also examine this choice empirically in \secref{sec:ablation} and find that using absolute time shifts instead of relative shifts results in much better performance.

We use a temporal resolution of 10 ms for our time events as some experiments have found this displacement to be approximately the limit of human perception \cite{friberg1993perception} (though others have reported smaller values e.g. 5 ms in Handel \cite{handel1993listening}). We leave open the possibility that our results could be improved further with finer event resolution, for example by predicting continuous times as in Kong et al. \cite{kong2020high}.

Decoding model output during inference is done with a simple greedy autoregressive algorithm.  We choose the maximum probability event at each step and feed that back into the network as the predicted event for that step.  We continue this process until the model predicts an EOS token.

Using an event sequence as our training target instead of piano roll matrices or other frame-based formats enables significant flexibility.  For example, we demonstrate in \secref{sec:ablation} that the exact same model configuration with the same inputs can be trained to predict only onsets (using just the Note, Time, and EOS events) or onsets, offsets, and velocities (using the full vocabulary above).  The only change required is using a different set of tokens as the training target.  This is in contrast to previous work where predicting a new feature required adding new output heads (or entire stacks), designing losses for those outputs, and modifying the (often non-differentiable) decoding algorithm to combine all model outputs into the final desired representation.

By using a sequence-to-sequence approach, our model can directly output our desired representation by jointly modeling audio features and language-like output dependencies in a fully differentiable, end-to-end training setting.  Adding new output features or changing the task definition is simply a matter of changing the tokens used to describe the target output.

\subsection{Sequence Length Considerations}

Transformers can attend to all tokens in a sequence at every layer, which is particularly suitable to a transcription task that requires fine grained information about pitch and timing for every event.  However, this attention mechanism comes at a space complexity of $O(n^2)$ with respect to sequence length $n$.  The practical consequence is that most audio sequences used for transcription cannot fit in memory.  To get around this problem, we split the audio sequence and its corresponding symbolic description into smaller segments during training and inference.

During training, we use the following procedure for every sequence in a batch:
\begin{enumerate}
\item Select a random audio segment from the full sequence for model input.  The length of the selected segment can vary from a single input frame to the maximum input length, and the starting position is selected from a uniform random distribution.

\item Select the symbolic segment for the training target that corresponds to the selected audio segment.  Because notes may start in one segment and end in another, the model is trained to be able to predict note-off events for cases where the note-on event was not observed.

\item Compute a spectrogram for the selected audio and map the symbolic sequence into our vocabulary (see \secref{subsec:inputsoutputs}).  Absolute time shifts within the symbolic segment are calculated such that time $0$ is the beginning of the segment.

\item Provide the continuous spectrogram input and one-hot-encoded MIDI-like events as a training example for the Transformer architecture.
\end{enumerate}

During inference, the following procedure is used:
\begin{enumerate}
\item Split the audio sequence into non-overlapping segments using the maximum input length when possible, and then compute spectrograms.

\item For each segment in turn, provide the spectrogram as input to the Transformer model and decode by greedily selecting the most likely token according to the model output at each step until an EOS token is predicted. Any tokens occurring after a time shift beyond the length of the audio segment will be discarded.

\item Concatenate the decoded events from all segments into a single sequence.  After concatenation, there may still be note-off events with no corresponding note-on; we remove these.  If we encounter a note-on event for a pitch that is already on, we end the note and start a new one.  At the end of the sequence, we end any active notes that are missing note-off events.

\end{enumerate}

The model is surprisingly capable at predicting note-on or note-off events where the corresponding event is in a different segment, as illustrated in \figref{fig:inference_frames}.  This ability is also empirically demonstrated by the model's results on the Onset, Offset, \& Velocity F1 scores in \secref{sec:evaluation}.

\section{Experiments}

\begin{table*}[t]
 \begin{center}
 \centering
\sisetup{table-format=2.2,round-mode=places,round-precision=2,table-number-alignment = center,detect-weight=true,detect-inline-weight=math}
 \begin{tabular}{ll|S|S|S}
  \toprule
  \multicolumn{2}{c|}{Model} & \multicolumn{1}{c|}{\parbox{2.25cm}{\centering Onset, Offset, \\ {\small \& } Velocity F1}} & \multicolumn{1}{c|}{\parbox{1.75cm}{\centering Onset \\ {\small \& } Offset F1}} & \multicolumn{1}{c}{\parbox{1.75cm}{\centering Onset \\ F1}} \\ \midrule
  \multirow{5}{*}{\rotatebox[origin=b]{90}{\parbox{1.5cm}{\centering MAESTRO \\ V1.0.0}}} &
  Transformer (ours) & \bfseries 82.18 & \bfseries 83.46 & 95.95 \\
  & Kong et al. 2020 \cite{kong2020high} & 80.92 & 82.47 & \bfseries 96.72 \\
  & Kwon et al. 2020 \cite{kwon2020polyphonic} & \multicolumn{1}{c|}{--} & 79.36 & 94.67 \\
  & Kim \& Bello 2019 \cite{kim2019adversarial} & 80.2 & 81.3 & 95.6 \\
  & Hawthorne et al. 2019 \cite{hawthorne2019enabling} & 77.54 & 80.50 & 95.32 \\
  \midrule
  \multirow{6}{*}{\rotatebox[origin=b]{90}{\parbox{1.5cm}{\centering MAESTRO \\ V3.0.0}}} & Transformer & \bfseries 82.75 & \bfseries 83.94 & 96.01 \\
  & Onsets only vocabulary & \multicolumn{1}{c|}{--} & \multicolumn{1}{c|}{--} & \bfseries 96.13 \\
  & STFT input & 81.81 & 82.92 & 95.44 \\
  & Raw samples input & 74.79 & 77.26 & 92.35 \\
  & Relative time shifts output & 66.25 & 67.35 & 80.02 \\
  & ``Base'' model size (100K steps) & 81.41 & 82.78 & 95.60 \\
  \bottomrule
 \end{tabular}
\end{center}
 \caption{MAESTRO test set results. Comparisons against previous work are done using V1.0.0 and comparisons against different model configurations are done using V3.0.0 because of its larger size.  All Transformer models were trained to 400K steps except for the ``Base'' configuration which was trained to 100K steps.}
 \label{tab:results}
\end{table*}

We trained our model with the Adafactor optimizer\cite{shazeer2018adafactor} using a batch size of 256, a constant learning rate of $1\mathrm{e}{-3}$, and dropout set to $.1$ for both sub-layer outputs and embedded inputs.  Batch size was selected to maximize training throughput because other batch sizes we tried during initial experimentation did not seem to make a difference in final performance.  The learning rate and dropout values were set to the same values used by T5 for fine-tuning tasks.

Input spectrograms were calculated using the Tensorflow\cite{tensorflow2015-whitepaper} \emph{tf.signal} library.  We used an audio sample rate of 16,000 kHz, an FFT length of 2048 samples, and a hop width of 128 samples.  We scaled the output to 512 mel bins (to match the model's embedding size) and used the log-scaled magnitude.

Input sequences were limited to 512 positions (511 spectrogram frames plus a learnable EOS embedding), and outputs were limited to 1024 positions (1023 symbolic tokens plus a learnable EOS embedding). This corresponds to a maximum segment length of $4.088$ seconds. We used 512 input positions to match the sequence length of T5, but future work could explore if other sequence lengths result in better performance. 1024 output positions were used because we found that 512 output positions were not always sufficient to symbolically describe the input audio.

We trained all models on 32 TPUv3 cores, resulting in a per-core batch size of 8.  We used this configuration for training speed, but the model is small enough to train on a single TPUv2 instance (8 cores).  Based on validation set results, overfitting did not seem to be a problem, so we allowed training to progress for 400K steps, which took about 2.5 days for our baseline models.

\subsection{Datasets}

To evaluate the performance of our model on the task of piano transcription, we use the MAESTRO dataset\cite{hawthorne2019enabling}, which contains about 200 hours of virtuosic piano performances captured with fine alignment between audio and ground truth note annotations.  For comparison against previous transcription work, we train on MAESTRO V1.0.0, but for other studies we use MAESTRO V3.0.0 because it contains an additional 92 performances containing 26 hours of data.  MAESTRO V3.0.0 also contains \textit{sostenuto} and \textit{una corda} pedal events, though our model (and evaluation) does not make use of these.  We also do not model sustain pedal events directly as in Kong et al. \cite{kong2020high}, instead extending note durations while the sustain pedal is pressed similar to Hawthorne et al. \cite{hawthorne2019enabling}.

\subsection{Evaluation}
\label{sec:evaluation}

In evaluating the performance of a piano transcription system, we use the Note F1 score metric: the harmonic mean of precision and recall in detecting individual notes.  This involves matching each predicted note with a unique ground truth note based on onset time, pitch, and optionally offset time.  Additionally, onset velocity can be used to discard matches with drastically different velocities.  We primarily use an F1 score that takes into account onsets, offsets, and velocities.  We also include results for F1 scores that consider only onsets or onsets and offsets.  We defer to the \emph{mir\_eval} \cite{raffel2014mir_eval} library for a precise definition of the (standard) transcription metrics we use.

Because piano is a percussive instrument, it is generally easier (and also more perceptually important) to accurately identify note onsets compared to offsets\cite{ycart2020investigating}.  We use \emph{mir\_eval}'s default match tolerance of 50 ms for onsets and the greater of 50 ms or 20\% of the note's duration for offsets.

\subsection{Comparison to Previous Work}

We compare our sequence-to-sequence approach with the reported scores from previous piano transcription papers on V1.0.0 of the MAESTRO dataset in \tabref{tab:results}.  Our method is able to achieve competitive F1 scores compared to the best existing approach while being conceptually quite simple, using a generic architecture and decoding algorithm and standard representations.

\subsection{Ablation Study}
\label{sec:ablation}

We perform an ablation study on some of the components of our model using V3.0.0 of the MAESTRO dataset in \tabref{tab:results}.  First, we verify the flexibility of this architecture to describe the input audio using a different set of features.  We modify the symbolic data to describe only onsets by using just the Note, Time, and EOS events.  The model trains successfully and achieves a high F1 score on this modified onset-only task.

Next, we investigate different input representations.  For ``STFT'', we remove the log mel scaling after FFT calculation.  This results in an input frame size of 1025, which is projected to the model embedding size of 512 by the dense layer.  For ``raw samples'', we simply split the audio samples into segments based on the hop width used for the spectrograms (128 samples) and use those directly as input, again projected to the embedding size by the dense layer.  Both of these configurations train successfully, but do not perform as well as log mel input.  We suspect this is because the mel scaling produces useful features that the model would otherwise have to use some of its capacity to extract.

We also verify that absolute time shifts are a better fit for this architecture by training a model with relative time shifts.  As expected, it does not perform as well.  Further, we noticed that the note-based evaluation metrics on the validation set varied dramatically during training, with sometimes as much as a 15 point difference in onset F1 score between adjacent validation steps. We hypothesize this is because small changes in relative time shift prediction are magnified when accumulated across a sequence to determine the absolute times needed for the metrics calculation; i.e., relative time shifts cause the resulting transcriptions to drift out of alignment with the audio.

Finally, we investigate if a larger model size would improve performance.  We scaled up our model size based on the ``base'' configuration from T5.  Specifically, we modified the following hyperparameters: $d_{\mathrm{model}} = 768$, $d_{\mathrm{ff}} = 2{,}048$, $12$ heads for attention, and $12$ layers each in the encoder and decoder.  These changes resulted in a model with 213M parameters, as opposed to our ``small'' configuration which had only 54M.  This model quickly overfit the training dataset, with scores on the validation set starting to decline after 100K steps, so we stopped training at that point.  Even with early stopping, this model does not perform as well as our ``small'' configuration, clearly demonstrating that even though piano transcription is a fairly complicated task, it does not require a particularly large Transformer.

\section{Conclusion and Future Work}

We have shown that a generic Transformer architecture trained to map spectrograms to MIDI-like output events with no pretraining can achieve state-of-the-art performance on automatic piano transcription.  We see this as an appeal to simplicity; we used standard formats and architectures as much as possible and were able to achieve results on par with models customized for piano transcription.  Possibly the main source of complexity in our setup is the splitting of examples into segments; future work could include the investigation of sparse attention mechanisms to enable the transcription of an entire piece of music in a single encoding and decoding pass. Also worth exploring would be the use of distillation\cite{hinton2015distilling} or related techniques to enable models like this to run in realtime on mobile devices or the web.

Our results suggest that a generic sequence-to-sequence framework with Transformers might also be beneficial for other MIR tasks, such as beat tracking, fundamental frequency estimation, chord estimation, etc. 
The field of Natural Language Processing has seen that a single large language model, such as GPT-3 or T5, has been capable of solving multiple tasks by leveraging the commonalities between tasks. 
We are excited by the possibility that similar phenomena could be possible with MIR tasks, and we hope that these results point toward possibilities for creating new MIR models by focusing on dataset creation and labeling rather than custom model design.

\bibliography{ISMIRtemplate}

\newpage
\onecolumn
\appendix

\section{Full Inference Results}

\begin{table*}[h]
\footnotesize
\centering
\sisetup{table-format=2.2,round-mode=places,round-precision=2,table-number-alignment = center,detect-weight=true,detect-inline-weight=math}
\begin{tabular}{cl|S|S|S|S|S|S|S|S|S}
 \toprule
 \multirow{2}{*}{MAESTRO} & \multirow{2}{*}{Model} & \multicolumn{3}{c|}{Onset, Offset, \& Velocity}
 & \multicolumn{3}{c|}{Onset \& Offset}
 & \multicolumn{3}{c}{Onset} \\ 
 
 & & \multicolumn{1}{c}{P} & \multicolumn{1}{c}{R} & \multicolumn{1}{c|}{F1} & 
 \multicolumn{1}{c}{P} & \multicolumn{1}{c}{R} & \multicolumn{1}{c|}{F1} & 
 \multicolumn{1}{c}{P} & \multicolumn{1}{c}{R} & \multicolumn{1}{c}{F1} \\
 \midrule
 
  V1.0.0 & Transformer & 84.45 & 80.07 & 82.18 & 85.77 & 81.31 & 83.46 & 98.62 & 93.46 & 95.95 \\
  
  \midrule
  
  \multirow{6}{*}{V3.0.0} & Transformer & \bfseries 84.95 & \bfseries 80.7 & \bfseries 82.75 & \bfseries 86.19 & \bfseries 81.86 & \bfseries 83.94 & \bfseries 98.61 & 93.6 & 96.01 \\
  & Onsets only vocabulary & {--} & {--} & {--} & {--} & {--} & {--} & 98.45 & \bfseries 93.97 & \bfseries 96.13 \\
  & STFT input & 84.47 & 79.38 & 81.81 & 85.63 & 80.45 & 82.92 & 98.59 & 92.57 & 95.44 \\
  & Raw samples input & 78.33 & 71.7 & 74.79 & 80.96 & 74.04 & 77.26 & 96.86 & 88.42 & 92.35 \\
  & Relative time shifts output & 68.33 & 64.34 & 66.25 & 69.48 & 65.4 & 67.35 & 82.57 & 77.68 & 80.02 \\
  & ``Base'' model size (100K steps) & 83.64 & 79.36 & 81.41 & 85.05 & 80.67 & 82.78 & 98.26 & 93.14 & 95.60 \\
  
 \bottomrule
 
\end{tabular}
 \caption{MAESTRO test set results for our models. All Transformer models were trained to 400K steps except for the ``Base'' configuration which was trained to 100K steps.}
 \label{tab:full-results}
\end{table*}

\section{Architecture Details}

\begin{table*}[h]
 \begin{center}
 \centering
\sisetup{table-format=1.1,round-mode=places,round-precision=1,table-number-alignment = center,detect-weight=true,detect-inline-weight=math}
 \begin{tabular}{ll}
  \toprule
  Parameter & Configuration \\ \midrule
  Embedding $d_{\mathrm{model}}$ & $512$ \\
  Feed-forward output $d_{\mathrm{ff}}$ & $1024$ \\
  Feed-forward activation & GEGLU \cite{shazeer2020glu} \\
  Key/value dimensionality $d_{\mathrm{kv}}$ & $64$ \\
  Attention heads & $6$ \\
  Encoder layers & $8$ \\
  Decoder layers & $8$ \\
  Input length & $512$ \\
  Output length & $1024$ \\
  Learning rate & $1\mathrm{e}{-3}$ (constant) \\
  Batch size & $256$ (total across devices) \\
  Dropout & $.1$ (sub-layer outputs and embedded inputs) \\
  Label smoothing & $0.0$ \\
  Optimizer & Adafactor\cite{shazeer2018adafactor} \\
  Activations & float32 \\
  Position encodings & Fixed absolute sinusoidal \cite{AttentionIsAllYouNeed} \\
  Decoding strategy & Autoregressive greedy argmax \\
  \bottomrule
 \end{tabular}
 \caption{The model configuration is based on the ``small'' model from T5 \cite{2020t5}, with modifications as suggested by the T5.1.1 recipe: \url{https://github.com/google-research/text-to-text-transfer-transformer/blob/master/released_checkpoints.md\#t511}.}
\end{center}
\end{table*}

\end{document}